\documentclass[preprint, superscriptaddress,amsmath,amssymb]{revtex4}
\usepackage{amstext}
\usepackage{amsmath}
\usepackage{amssymb}
\usepackage{graphicx}
\usepackage{latexsym}
\usepackage{bm}
\date{\today}

\begin{document}
\title{Quantum phase estimation algorithms with delays: effects of dynamical phases}
\author{L.F. Wei}
\affiliation{Frontier Research System, The Institute of Physical
and Chemical Research (RIKEN), Wako-shi, Saitama, 351-0198, Japan}
\affiliation{Institute of Quantum Optics and Quantum Information,
Department of Physics, Shanghai Jiaotong University, Shanghai
200030, P.R. China }
\author{Franco Nori}
\affiliation{Frontier Research System, The Institute of Physical
and Chemical Research (RIKEN), Wako-shi, Saitama, 351-0198, Japan}
\affiliation{Center of Theoretical Physics, Physics Department,
Center for the Study of Complex Systems, University of Michigan,
Ann Arbor, Michigan 48109-1120, USA}
\thanks{Permanent address}

\begin{abstract}

The unavoidable finite time intervals between the sequential
operations needed for performing practical quantum computing can
degrade the performance of quantum computers. During these delays,
unwanted relative dynamical phases are produced due to the free
evolution of the superposition wave-function of the qubits. In
general, these coherent ``errors" modify the desired quantum
interferences and thus spoil the correct results, compared to the
ideal standard quantum computing that does not consider the
effects of delays between successive unitary operations. Here, we
show that, in the framework of the quantum phase estimation
algorithm, these coherent phase ``errors", produced by the time
delays between sequential operations, can be avoided by setting up
the delay times to satisfy certain matching conditions.

PACS number(s): 03.67.Lx

\end{abstract}

\maketitle

\section{Introduction}

Building a prototype quantum information processor has attracted
considerable interest during the past decade (see, e.g.,
\cite{Nielsen00}). This desired device should be able to
simultaneously accept many different possible inputs and
subsequently evolve them into a corresponding quantum mechanical
superposition of outputs. The proposed quantum algorithms are
usually constructed for ideal quantum computers. In reality, any
physical realization of such a computing process must treat
various errors arising from various noise and imperfections (see,
e.g., \cite{Nielsen00,Mussinger00,Long}). Physically, these errors
can be distinguished into two different kinds: incoherent and
coherent errors. The incoherent perturbations, originating from
the coupling of the quantum computer to an uncontrollable external
environment, result in decoherence and stochastic errors. Coherent
errors usually arise from non-ideal quantum gates which lead to a
unitary but non-ideal temporal evolution of the quantum algorithm.
So far, almost all previous works (see, e.g., \cite{Shor95,
Steane98, Lidar98, Zanardi97}) have been concerned with quantum
errors arising from the decoherence due to interactions with the
external environment and external operational imperfections. Here,
we will not be concerned with these two types of
externally-induced errors, but will focus instead on intrinsic
ones. The coherent errors we consider here relate to the intrinsic
dynamical evolution of the qubits between operations. This has not
been paid much attention until a recent work in \cite{Berman00},
where a kind of dynamical phase error was introduced. It is
well-known that a practical quantum computing process usually
consists of a number of sequential quantum unitary operations.
These transformations operate on superposition states and evolve
the quantum register from the initial states (input) into the
desired final states (output). According to the Schr\"odinger
equation, the superposition wave function oscillates fast during
the finite-time delay between two sequential operations. In
general, these oscillations modify the desired quantum
interferences and thus spoil the correct computational results,
expected by the ideal quantum algorithms without any operational
delay.

Two different strategies have been proposed to deal with these
coherent errors. One is the so-called ``avoiding error" approach
proposed by Makhlin {\it et al} in Ref. \cite{MSS99}. Its key idea
is to let the Hamiltonian of the bare two-level physical system be
zero by properly setting up experimental parameters. Thus the
system does not evolve during the delays. This requirement is
restrictive and cannot be easily implemented for some physical
setups of quantum computing e.g., for trapped ions. A modified
approach to remove this stringent condition was proposed by Feng
in Ref. \cite{Feng01}, where a pair of degenerate quantum states
of a pair of two-level systems are used to encode two logic states
of a single qubit. During the delay these logical states acquire a
common dynamical phase, which is the global phase without any
physical meaning. Thus the above dynamical error can be avoided
efficiently. However, this modified scheme complicates the process
of encoding information. Another strategy to this problem was
proposed by Berman {\it et al.\/} \cite{Berman00}. They pointed
out that the unwanted dynamical oscillations can be routinely
eliminated by introducing a ``natural" phase, which can be induced
by using a stable continuous reference oscillation for each
quantum transition in the computing process. However, this scheme
only does well for the resonant implementations of quantum
computation. The additional reference pulses also complicate the
quantum computing process and may result in other operational
errors.

We show in this paper that, in the framework of the quantum phase
estimation algorithm, the coherent phase errors, produced by the
free evolutions of the superposition wave functions of bare
two-level systems, can be avoided simply and effectively by
setting up the delay time intervals appropriately. The proposed
matching condition can be considered a sort of strobed operation
(with strobe frequencies corresponding to each different
transition energy). For simplicity, we simplify each quantum
algorithm to a three-step functional process, namely: preparation,
evolution, and measurement. All the functional operations in this
three-step process are assumed to be carried out in an
infinitesimally short time duration, and thus only the delays
between them, instead of the operations themselves, are
considered. The effects of the environment decoherence and the
operational imperfections are neglected in the present treatment.

The paper is organized as follows. In Section II, we present our
general approach with the phase estimation algorithm. Sec. III
gives a few special demonstrations and shows how to perform
quantum order-finding and quantum counting algorithms in the
presence of operational delays. Finally, we give a short summary
and discussion in Sec. IV.

\section{Phase estimation algorithm with operational delays}

Our discussion begins with the phase estimation algorithm
\cite{Kitaev95,Cleve98} and its finite-time implementation with
some delays. The programs for some of the existing other important
quantum algorithms, such as quantum factoring and counting ones,
can be reformulated in terms of this problem.
The goal of the phase estimation algorithm is to obtain an $n-$bit
estimation of the eigenvalue $\exp(i\phi)$ of a unitary operation
$\hat{U}_T$,
\begin{equation}
\hat{U}_T\,|\phi\rangle_T=e^{i\phi}\,|\phi\rangle_T,
\end{equation}
if the corresponding eigenvector $|\phi\rangle_T$, and the devices
that can perform operations $\hat{U}_T$, $\hat{U}^2_T$,
$\hat{U}_T^4,\cdots$, and $\hat{U}^{2^n}_T$, are given initially.
Two quantum registers are required to perform this algorithm. One
is the target register, whose quantum state is kept in the
eigenstate $|\phi\rangle_T$ of the unitary operator $\hat{U}_T$.
Another one, with $n$ physical qubits and called the {\it index}
register, is used to read the corresponding estimation results.
The needed number of qubits $n$ in the {\it index} register
depends on the desired accuracy and on the success probability of
the algorithm.
The most direct application \cite{Abrams99} of this algorithm is
to find eigenvalues and eigenvectors of a local Hamiltonian
$\hat{H}_T$
by determining the time-evolution unitary operator
$\hat{U}_T=\exp(-i\hat{H}_Tt/\hbar)$.
The phase estimation algorithm can be viewed as a quantum
nondemolition measurement, and can also be used to generate
eigenstates of the corresponding unitary operator $\hat{U}_T$
\cite{Travaglione02}.

The ideal quantum algorithm usually assumes that the quantum
computing process can be continuously performed by using a series
of sequential operations without any time-delay between them.
In reality, a delay between two sequential operations always
exists, introducing errors that need to be corrected.
For simplicity, we reduce the phase estimation algorithm to a
three-step functional process, namely: initialization, global
phase shift, and measurement. All the functional operations in
this three-step process are assumed to be carried out exactly, and
thus only the delays between them, instead of the operations
themselves, are considered.
Such a simplified finite-time implementation of the phase
estimation algorithm is sketched in Fig.~1. For convenience we
distinguish the physical qubit and the logic qubit in the index
register.
The physical qubit is just a two-level physical system and the
logical qubit is the unit of binary information.
Unlike the scheme in \cite{Feng01}, wherein two physical qubits
are used to encode one logical qubit, in the present work {\it
one} physical qubit is enough to encode {\it one} logical qubit.
The symbol $|a_j\rangle_k$ with $a=0,1,\,\,j,k=0,1,...,n-1$ means
that the $k$th logical qubit is encoded by the $j$th physical
qubit.
$|a_j\rangle$ is the eigenstate of the bare Hamiltonian of the
$j$th physical qubit corresponding to the eigenvalue $E_a$.
\begin{figure}
\vspace{2.5cm}
\hspace{2cm}\includegraphics[width=18.8cm]{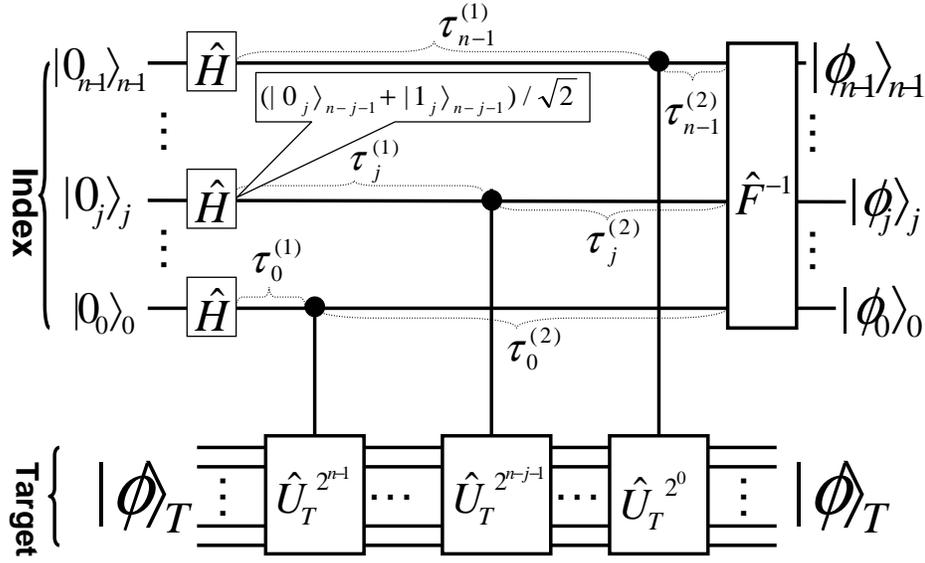}
\vspace{-10cm} \caption{Quantum phase estimation with operational
delays. Note that: 1) there is an operational delay $\tau_j^{(m)}
(m=1,2)$ between successive quantum operations on the $j$th
physical qubit, and 2) the $j$th logical qubit is changed to the
$(n-j-1)$th one after the Hadamard gate $\hat{H}$ and inverse QFT
$\hat{F}^{-1}$. Here, $\tau^{(1)}_j$ is time delay between the
$\hat{H}$ and $\hat{U}_T^{2^{n-j-1}}$ operations, while
$\tau^{(2)}_j$ is the delay between $\hat{U}_T^{2^{n-j-1}}$ and
$\hat{F}^{-1}$.}
\end{figure}

The quantum phase estimation algorithm with operational delays can
be divided into three distinct functional steps:

\subsection{Initialization}
First, we initialize the index register with $n$ physical qubits
in an equal-weight superposition of all logical states.
This can be performed by applying the Hadamard transform to its
ground state $ |0\rangle_I=\prod_{j=n-1}^0|0\rangle_j. $
Note that the target register holds an eigenstate $|\phi\rangle_T$
of $\hat{U}_T$ with eigenvalue $\exp(i\phi)$.
Hereafter, the sub-index $I$ will denote the index state, while
the subindex $T$ refers to the target state.
The computational initial state of the whole system is
\begin{eqnarray}
|\Psi(0)\rangle=\left\{\prod_{j=n-1}^{0}\hat{H}_j\,|0_j\rangle_j\right\}_I\otimes
               |\phi\rangle_T
               =\frac{1}{\sqrt{2^n}}\sum_{k=0}^{2^n-1}|k\rangle_I\otimes
               |\phi\rangle_T,\,\,\,\hat{H}_j=\frac{1}{\sqrt{2}} \left(
\begin{array}{cc}
1&1\\
1&-1
\end{array}
\right)_j,
\end{eqnarray}
where $|k\rangle_I=|a_{0}\rangle^k_{n-1}\otimes\cdots\otimes
|a_{n-1}\rangle^k_0$
are the number states of the index register, and $\hat{H}_j$ is
the Hadamard transform applied to the $j$th logical qubit.
For convenience, in this paper the $j$th logical qubit is changed
into the $(n-1-j)$th logical qubit when applying either the
Hadamard or the (inverse) quantum Fourier transform (QFT).
Of course, the order of the physical qubits is not changed.

After a finite time delay $\tau^{(1)}_j$ for the $j$th physical
qubit, the initial state $|\Psi(0)\rangle$ of the whole system
evolves to
\begin{equation}
|\Phi\{\tau^{(1)}_j\}\rangle=\left\{\prod_{j=n-1}^{0}\frac{1}{\sqrt{2}}
\left(e^{-iE^0_j\tau^{(1)}_j}|0_j\rangle_{n-j-1}\,+\,e^{-iE^1_j\tau^{(1)}_j}|1_j\rangle_{n-j-1}\right)\right\}_I\otimes
               |\phi\rangle_T,
\end{equation}
with $E^0_j$ and $E^1_j$ being the eigenvalues of the Hamiltonian
for the $j$th bare physical qubit corresponding to the
eigenvectors $|0_j\rangle$ and $|1_j\rangle$, respectively.

\subsection{Global phase shift}

Second, we shift the ``global" phase in the eigenvector of the
operator $\hat{U}$ into a measurable relative phase.
This can be achieved by using the ``phase kick-back" technique
\cite{Cleve98}.
Indeed, after applying a controlled-$\hat{U}^{2^j}_T$ operation
$c-\hat{U}_j$, defined by
\begin{equation}
c-\hat{U}_j\,=\,|1\rangle_{j}\,_{j}\langle
1|\otimes\hat{U}_T^{2^j}\,+\,|0\rangle_{j}\,_{j}\langle
0|\otimes\hat{I}_T,
\end{equation}
to the $j$th logical qubit, the state
$|\Phi\{\tau_j^{(1)}\}\rangle$ is evolved to
\begin{eqnarray}
|\Psi\{\tau^{(1)}_j\}\rangle&=&\prod_{j=n-1}^0\left(c-\hat{U}_j\right)|\Phi\{\tau^{(1)}_j\}\rangle\nonumber\\
&=&\frac{1}{\sqrt{2}}
\left(e^{-iE^0_0\tau^{(1)}_0}|0_0\rangle_{n-1}
+e^{-iE^1_0\tau^{(1)}_0}e^{i2^{n-1}\phi}|1_0\rangle_{n-1}\right)\otimes\nonumber\\
&&\cdots\otimes \frac{1}{\sqrt{2}}
\left(e^{-iE^0_{n-1}\tau^{(1)}_{n-1}}|0_{n-1}\rangle_{0}
+e^{-iE^1_{n-1}\tau^{(1)}_{n-1}}e^{i2^{0}\phi}|1_{n-1}\rangle_{0}\right)\otimes
|\phi\rangle_T.
\end{eqnarray}
Here $|1\rangle_{j}\,_{j}\langle 1|$ and
$|0\rangle_{j}\,_{j}\langle 0|$ are the projectors of the $j$th
logical qubit.
$\hat{I}_T$ is the identity or unity operation.
The controlled-$\hat{U}^{2^j}_T$ operator means that, if the $j$th
logical qubit in the index register is in the state $|1\rangle_j$,
the $2^j$-fold iteration of $\hat{U}_T$ is applied to the target
register.
The ``global" phase in the eigenvector of the operator
$\hat{U}^{2^j}_T$ is changed as the measurable relative phases in
the states of the index qubits.

Before the next step in the operation of the algorithm there is
another finite-time delay $\tau^{(2)}_j$ for the $j$th physical
qubit.
During this time interval each physical qubit of the index
register evolves again freely according to the Schr\"odinger
equation,
while the target register is assumed to be still in the state
$|\phi\rangle_T$.
As a consequence, the state of the whole system becomes
\begin{eqnarray}
|\Phi\{\tau_j\}\rangle&=&\frac{1}{\sqrt{2}}
\left(e^{-iE^0_0\tau_0}\,|0_0\rangle_{n-1}
+e^{-iE^1_0\tau_0}e^{i2^{n-1}\phi}|1_0\rangle_{n-1}\right)\otimes
\nonumber\\
&&\cdots\otimes \frac{1}{\sqrt{2}}
\left(e^{-iE^0_{n-1}\tau_{n-1}}|0_{n-1}\rangle_{0}
+e^{-iE^1_{n-1}\tau_{n-1}}e^{i2^{0}\phi}|1_{n-1}\rangle_{0}\right)\otimes
|\phi\rangle_T,
\end{eqnarray}
with
\begin{equation}
\tau_j=\tau^{(1)}_j+\tau^{(2)}_j
\end{equation}
being the total delay before and after the
controlled-$\hat{U}^{2^{n-j-1}}_T$ operation. Note that the
dynamical phases of the index qubit can be add up before and after
this, as controlled-$\hat{U}^{2^{n-j-1}}_T$ operator is diagonal
in the basis of the $(n-j-1)$th logical qubit of the index
register.

\subsection{Measurement}

Third, we finally apply the inverse quantum Fourier transform
(QFT) on the index register to measure the phase in the
eigenvector of the unitary operator $\hat{U}_T$.
The inverse QFT, defined by the formula
\begin{equation}
QFT^{-1}:
|k\rangle\,\,\longrightarrow\,\hat{F}^{-1}|k\rangle=\frac{1}{\sqrt{2^n}}
\sum_{y=0}^{2^n-1}\exp\left(-2\pi i\frac{k\cdot
l}{2^n}\right)\,|l\rangle,
\end{equation}
can be performed by using the sequential unitary operations $
\hat{F}^{-1}$$\,=\,$$\hat{F}^{\dagger}$$\,
=\hat{H}_0\hat{R}_{0,1}^{\dagger}
\cdots\hat{H}_{n-2}\cdots\hat{R}_{0,n-1}^{\dagger}\cdots
\hat{R}_{n-2,n-1}^{\dagger}\hat{H}_{n-1}, $
to the corresponding logical qubits.
Here,
$$
\hat{R}^{\dagger}_{j-k,j}= \left(
\begin{array}{cccc}
1&0&0&0\\
0&1&0&0\\
0&0&1&0\\
0&0&0&e^{-i\pi/2^k}
\end{array}
\right)_{j-k,\,j},
$$
is a two-qubit controlled-phase operation.
It implies that the state $|1\rangle_{j}$ of the target $j$th
logical qubit will change by a phase $\exp(-i\pi/2^k)$, if the
control $(j-k)$th logical qubit is in the state $|1\rangle_{j-k}$.
If the phase $\phi$ can be exactly written as a $n$-bit binary
expansion, i.e.,
\begin{equation}
\phi=2\pi(.\phi_0\cdots
\phi_{n-1})=\frac{\phi_0}{2^n}+\frac{\phi_1}{2^{n-1}}+\cdots+\frac{\phi_{n-1}}{2},\,\,\,\,\phi_j=0,1,\,\,\,\,j=0,1,...,n-1,
\end{equation}
then the expected final output state of the index register, after
applying the inverse QFT,  is the following product state
\begin{equation}
|\Psi\{\tau_j\}\rangle_I=|\phi_{n-1}\rangle_{n-1}\otimes\cdots
|\phi_j\rangle_j\cdots\otimes |\phi_0\rangle_0.
\end{equation}
However, the existing dynamical phase error, arising from the free
evolution of the physical qubits during the delays, may spoil the
desired results.
For example, measuring the $j$th physical qubit in the
computational basis $\{|0\rangle,|1\rangle\}$, we have
\begin{eqnarray}
\hat{F}^{\dagger}: \left[e^{-iE^0_j\tau_j}|0_j\rangle_{n-1-j}
+e^{-iE^1_j\tau_j}\,e^{i2^{n-1}2\pi(.\phi_0\cdots\phi_{n-1-j})}|1_j\rangle_{n-1-j}\right]/\sqrt{2}
\longrightarrow \nonumber\\
e^{-iE^0_j\tau_j}
\left[\left(1+e^{-i\Delta_j
\tau_j}\,e^{i\pi\phi_j}\right)\,|0_j\rangle_{j}
+\left(1-e^{-i\Delta_j\tau_j}\,e^{i\pi\phi_j}\right)\,|1_j\rangle_{j}\right]/\sqrt{2}.
\end{eqnarray}
The expected result $|\phi_j\rangle_j$ is obtained with the
following probability
\begin{equation}
P_{\phi_j}=\frac{1}{2}\left[1+\cos(\Delta_j\tau_j)\right],\,\,\,\,\Delta_j=E^1_j-E^0_j,
\end{equation}
while an error output state $|\phi_j\oplus 1\rangle_j$ is obtained
with the probability $P_{\phi_j\oplus
1}=\left[1-\cos(\Delta_j\tau_j)\right]/2$. Here $\oplus$ refers to
addition modulo two. Note that the above probability (12) of
obtaining the corret result only depends on the {\it total} delay
time $\tau_j$, but {\it not} directly on the individual time
internals $\tau_j^{(m)};\;m=1,2$.

Obviously, if $\tau^{(1)}=\tau^{(2)}=0$, i.e., for the ideal
algorithm realization without any delay, one obtains the desired
output $|\phi_j\rangle_j$.
While for the realistic case where $\tau_j^{(1)},\tau_j^{(2)}\neq
0$, the required quantum inference may be modified, and thus the
real output may not be the expected one.
A worst case scenario is produced if
\begin{equation}
\Delta_j\,\tau_j=(2\,l+1)\pi,\,\,\,l=0,1,2,...,
\end{equation}
because the corresponding error-state output is $|\phi_j\oplus
1\rangle_j$, which is incorrect.
However, if the following matching condition
\begin{equation}
\Delta_j\,\tau_j=2\,(l+1)\pi,
\end{equation}
is satisfied, one obtains the desired output $|\phi_j\rangle_j$,
and thus the fast oscillation of the superpositional wave function
is suppressed in the output of the computation.
Above, $ \tau_j=\tau_j^{(1)}+\tau_j^{(2)} $ is the {\it total
effective delay time} of the $j$th physical qubit in the
algorithm.
The condition in Eq.~(14) is desirable for implementing quantum
algorithms with an arbitrary number of qubits and includes as a
particular case, the less general condition in \cite{Berman00} for
the finite-time implementation of the 4-qubit Shor's algorithm.
\section{Example and applications}

We now demonstrate the above general approach via a simple
example, and show the effects of dynamical phases in finite-time
implementations of a few quantum algorithms.
\subsection{NOT gate eigenvalue}

First, we wish to determine the eigenvalue of the Pauli operator
$\hat{\sigma}_x$, or NOT gate,
by running the realistic single-qubit phase estimation algorithm
discussed above.
Assuming that the single-qubit target register is prepared into
one of the eigenstates
\begin{equation}
|\phi\rangle_T\,=\,|\pm\rangle_T\,=\,\frac{1}{\sqrt{2}}\left(
\begin{array}{c}
1\\
\pm 1
\end{array}
\right)_T,
\end{equation}
corresponding to the eigenvalues $e^{i\phi}$ with $\phi=0,\,\pi$,
respectively.
According to the above discussions, the final state of the index
single-qubit register, after the single-qubit measurement just
performed by Hadamard transform, can be written as
\begin{equation}
|\Psi(\tau)\rangle_I=\frac{1}{2}\left\{\left[\,1+e^{-i\Delta\tau+i\phi}\right]
e^{-iE^0\tau}\,|0\rangle_I
+\left[\,1-e^{-i\Delta\tau+i\phi}\right]e^{-iE^1\tau}\,|1\rangle_I\right\}.
\end{equation}
This implies that the probability for the index register to be
finally in the state $|0\rangle_I$ or $|1\rangle_I$ is
\begin{equation}
P_0(\tau)=\frac{1}{2}\left[1+\cos\phi\cos(\Delta\tau)+\sin\phi\sin(\Delta\tau)\right],
\end{equation}
or
\begin{equation}
P_1(\tau)=\frac{1}{2}\left[1-\cos\phi\cos(\Delta\tau)+\sin\phi\sin(\Delta\tau)\right].
\end{equation}
If the target register is in the eigenstate $|+\rangle_I$ of
operator $\hat{\sigma}_x$ with eigenvalue $+1$,\,i.e.,\,$\phi=0$,
the probability of getting the expected output $|0\rangle_I$ is
$P_0(\tau)=1$, if the condition (14) is satisfied.
However, if the condition (13) is satisfied, the index register
will show the error output, i.e., $|1\rangle_I$.

\subsection{Dynamical phase effects in the quantum order-finding algorithm with delays}

Shor's algorithm \cite{Shor94} for factoring a given number $N$ is
based on calculating the period of the function $f(x)=y^x\, {\rm
mod}\, N$ for a randomly selected integer $y$ between $1$ and $N$.
Once, the order $r$ of $y\, {\rm mod}\, N$ is known, factors of
$N$ are obtained by calculating the greatest common divisor of $N$
and $y^{r/2}\pm 1$.
A finite-time implementation of the order-finding algorithm can be
translated to the above quantum phase estimation algorithm with
delays.
Here, the unitary operator whose eigenvalue we want to estimate is
the unitary transformation $\hat{U}_y$, with
$\hat{U}_y^r=\hat{I}$, which maps $|x\rangle$ to $|yx\, {\rm
mod}\, N\rangle$ and
\begin{equation}
\hat{U}_y|u_k\rangle=\exp\left(i\frac{2\pi
k}{r}\right)|u_k\rangle,\,
|u_k\rangle=\frac{1}{\sqrt{r}}\sum_{x=0}^{r-1}\exp\left(\frac{2\pi
ikx}{r}\right)|y^x\,{\rm mod}\,N\rangle, \,k=0,...r-1.
\end{equation}
By the phase estimation algorithm, we can measure the eigenvalue
$\exp(2\pi ik/r)$ and consequently get the order $r$.
However, the present target register can not be prepared
accurately in one of the eigenvectors $|u_k\rangle$, as the order
$r$ is initially unknown.
It is noted that $ \sum_{k=0}^{r-1}|u_k\rangle/\sqrt{r}=|1\rangle,
$ and $|1\rangle$ is an easy state to prepare.
Thus, the algorithm may be run by initially generating a
superposition of all eigenstates of the operator $\hat{U}_y$,
rather than one of them accurately.

Without loss of generality, we demonstrate our discussion with the
simplest meaningful instance of Shor's algorithm, i.e., the
factorization of $N=15$ with $y=7$,
which had been implemented in a recent NMR experiment
\cite{Lieven01}.
In this simplest case, the order $r$ is the power of two, i.e.,
$r=2^n,\,n=2$, and thus
the expected phase estimation algorithm can measure exactly the
$n$-qubit eigenvalue $k/2^n:\,\,\, k=\sum_{j=0}^{n-1}k_j\,
2^j,\,\,k_j=0,1$.
From the measurement eigenvalues we can obtain the order $r$ by
checking if $y^r\,{\rm mod}\,N=1$.
Following the corresponding experimental
demonstration~\cite{Lieven01}, we need an index register with
$n=2$ physical qubits to measure the eigenvalues of the present
unitary operator $\hat{U}_y$, and a target register with $m=4$
physical qubits to represent the state
$|1\rangle_T=\sum_{k=0}^3|u_k\rangle_T/2,\,|u_k\rangle_T=\sum_{x=0}^3\,\exp\left(-2\pi
ikx/2^2\right)|7^{x}\,{\rm mod}\,15\rangle_T/2$,
which, in fact, is the equal-weight superposition of all the
eigenvectors of the operator $\hat{U}_y:
|x\rangle_T\longrightarrow |7^{\,x}\,{\rm mod}\,15\rangle_T,\,
x=0,1,2,3$,
with
$\hat{U}_y\,|u_k\rangle_T=\exp\left(2\pi
ik/2^2\right)\,|u_k\rangle_T$.
According to the three-step finite-time implementation of the
phase estimation discussed in the last section, one can easily
prove that the whole system is in the following entangled state
\begin{eqnarray}
|\Phi\{\tau_j\}\rangle=
\frac{1}{2}\sum_{k=0}^3
\prod_{j=1}^0\left\{
\frac{1}{\sqrt{2}}\left[|0_j\rangle_{1-j}+\exp\left(-i\Delta_j\tau_j+\frac{2\pi
i2^{(1-j)}k}{2^2}\right)|1_j\rangle_{1-j}\right)\right\}_I \otimes
|u_k\rangle_T,
\end{eqnarray}
before the index register is measured by using the inverse QFT.
Here, the unimportant global dynamical phase factor
$\exp(-2iE^0_j\tau_j)$ is neglected.

In the ideal case, i.e., $\tau^{(1)}_j=\tau^{(2)}_j=0$, measuring
the index register by the inverse QFT will, with a probability
equal to $1/4$, produce the expected output state
\begin{equation}
|\Psi_{\rm out}\rangle_I=|k_1\rangle_1\,\otimes |k_0\rangle_0.
\end{equation}
Simultaneously, the target register will ``collapse" into the
state of the corresponding expected eigenvector $|u_k\rangle$.
Once a measurement output, i.e., $k/2^2=(2^1k_1+2^0k_0)/2^2$ is
known,
the order is efficiently verified by checking if $y^i\,{\rm
mod}\,N=1$ for $i=2^2/k,\,2\cdot 2^2/k,...,r$.
For example, if the output is $k=3$, i.e., $|\Psi_{\rm
out}\rangle_I=|1_1\rangle_1\otimes |1_0\rangle_0$,
the order can be verified by testing $y^i\,{\rm mod}\,N=1$ for
$i=\{2^2/3,\,2\cdot 2^2/3,\,3\cdot 2^2/3=4=r\}$.
Of course, the algorithm fails if the output is $k=0$, i.e., the
target register collapses into the corresponding eigenvector
$|u_0\rangle$.
However, these deductions may be modified in a realistic quantum
computing process where the delays exist, i.e.,
$\tau_j^{(1)},\tau_j^{(2)}\neq 0$.
In fact, one can easily see from Eq.~(18) that, after applying the
inverse QFT, if the target register collapses into the state
$|u_k\rangle$,
the output in the index register reads
\begin{eqnarray}
|\Psi_{\rm
out}\rangle_I=\prod_{j=1}^0\left[\frac{1}{2}\left(1+e^{-i\Delta_j\tau_j+\pi
ik_j}\right)|0_j\rangle_j
+\frac{1}{2}\left(1-e^{-i\Delta_j\tau_j+\pi
ik_j}\right)|1_j\rangle_j\right].
\end{eqnarray}
Therefore, the expected state $|k_1\rangle_1\otimes|k_0\rangle_0$
is obtained, only if the delays are set up to satisfy the matching
condition (14). Otherwise, some errors may appear in the index
register.
In particular, an undesirable bit flip error will be produced if
Eq.~(13) is satisfied. For example, if the target register
collapses into the state $|u_3\rangle_T$, the index register
generates a null $|0\rangle_I=|0_1\rangle_1\otimes |0_0\rangle_0$,
but not the expected output $|3\rangle_I=|1_1\rangle_1\otimes
|1_0\rangle_0$.

\subsection{Quantum counting algorithm with operational delays}

Quantum counting is an application of the phase estimation
procedure to estimate the eigenvalues of the Grover iteration
\cite{Jones99,Grover97},
\begin{equation}
\hat{G}\,=\,-\hat{A}\hat{U}_0\hat{A}^{-1}\hat{U}_f.
\end{equation}
Here, $\hat{A}$ is any operator which maps $|0\rangle$ to
$\sum_{x=0}^{N-1}|x\rangle/\sqrt{N}$, $\hat{U}_0$ maps $|0\rangle$
to $-|0\rangle$ and $\hat{U}_f$ maps $|x\rangle$ to
$(-1)^{f(x)}|x\rangle$.
This algorithm enables us to estimate the number of solutions to
the search problem, as the Grover iterate is almost periodic with
a period dependent on the number of solutions.
Indeed, from the following equation
\begin{equation}
\hat{G}|\Psi_{\pm}\rangle=\exp(\pm 2\pi
i\omega_l)|\Psi_{\pm}\rangle,\,\,\,l=0,1,2,...,N,
\end{equation}
with $ |\Psi_{\pm}\rangle=\left(|X_1\rangle\,\pm\,
i|X_0\rangle\right)/\sqrt{2},\,\,\,\,\exp(\pm 2\pi
i\omega_l)=1-2l/N\pm 2i\sqrt{l/N-\left(l/N\right)^2}, $ and $
|X_1\rangle=\sum_{f(x)=1}|x\rangle/\sqrt{l},\,
|X_0\rangle=\sum_{f(x)=0}|x\rangle/\sqrt{N-l},$
we see that either $\omega_l$ or $-\omega_l$ can be estimated by
using the phase estimation algorithm.
This gives us an estimation of $l$, the number of solutions.

In order to explicitly demonstrate how the dynamical phase error
reveals in quantum counting, we consider the simple case where
$l=N/4$. The expected eigenvalues we want to estimate are
$\exp(\pm \pi i/3)$, corresponding to the target register being
kept in the eigenstates $ |\Psi_{\pm}\rangle$.
However, in this case the expected output $\omega_1=1/6$ cannot be
expressed exactly in a $n$-bit expansion.
Following Jones {\it et al.} \cite{Jones99} and Lee {\it et al.}
\cite{Lee02}, we now adopt the ensemble measurement to
approximately characterize the final state of the index register.
The algorithm operates on two registers: a single-qubit index
register and the target register with $m$ qubits, which are
initially prepared in their ground state:
$|\psi(0)\rangle_I=|0\rangle$,\,
$|\psi(0)\rangle_T=\prod_{j=m-1}^0|0\rangle_j$. A quantum counting
algorithm with delays can also be performed by three operational
steps:

1) Applying the Hadamard transform to two registers
simultaneously, we have
\begin{equation}
|\Psi_1\rangle=|\psi_1\rangle_I\otimes |\psi_1\rangle_T,
\end{equation}
with $
|\psi_1\rangle_I=\hat{H}|0\rangle_I=\left(|0\rangle+|1\rangle\right)/\sqrt{2},
$ and $
|\psi_1\rangle_T=c_{+}|\Psi_{+}\rangle_T+c_{-}|\Psi_{-}\rangle_T,\,
c_{\pm}=\mp i\exp(\pm i\pi/6)/\sqrt{2}. $

2) After the first finite-time delay $\tau^{(1)}$, we apply the
controlled operation $ c-\hat{G}=|1\rangle_I\,_I\langle
1|\otimes\hat{G}_T+|0\rangle_I\,_I\langle 0|\otimes\hat{I}_T $ to
the state $|\Psi_1\rangle$, and have
\begin{equation}
|\Psi_2\rangle=\sum_{j=\pm}\frac{c_j}{\sqrt{2}}\left[|0\rangle_I+e^{i(2\pi
j\omega_l-\Delta\tau^{(1)})}|1\rangle_I\right]\otimes\,|\Psi_j\rangle_T,
\end{equation}
After $k$ repetitions of the above operations, the state of the
system becomes
\begin{equation}
|\Psi_3\rangle\,=\sum_{j=\pm}\frac{c_j}{\sqrt{2}}\left[|0\rangle_I+e^{i(2\pi
jk\omega_l-\Delta(\tau^{(1)}+\tau^{(2)}+...+\tau^{(k-1)})}|1\rangle_I\right]\otimes\,|\Psi_j\rangle_T.
\end{equation}
Above, the controlled operation $c-\hat{G}$ means that the
operation $\hat{G}$ is applied to the target register only when
the control qubit is in state $|1\rangle_I$.

3) After another finite-time delay $\tau^{(k)}$, we apply a second
Hadamard transform to the control qubit, producing
\begin{equation}
|\Psi_4\rangle=\frac{1}{2}\sum_{j=\pm}c_j\left[\left(1+e^{i(2\pi
kj\omega_l-\Delta\tau)}\right)\,|0\rangle_I+\left(1-e^{i(2\pi
kj\omega_l-\Delta\tau)}\right)\,|1\rangle_I\right]\otimes|\Psi_j\rangle_T,\,\,
\tau=\sum_{m=1}^k\tau^{(m)},
\end{equation}
and then the expectation value of $\hat{\sigma}_z$ is measured to
characterize the final state of the index register.
This corresponds to determining the population difference between
$|0\rangle_{I}\,_{I}\langle 0|$ and $|1\rangle_{I}\,_{I}\langle
1|$ in the state $|\Psi_4)\rangle$,
and the result can be expressed as
\begin{equation}
<\hat{\sigma}_z>_I\,=\,\cos(2\pi k\omega_l-\Delta\tau).
\end{equation}
The expected result for the ideal case, i.e., $\tau^{(m)}=0$, is
$<\hat{\sigma}_z>_I(\tau)=\cos(2\pi k\omega_l)$, and the value
$\omega_l$ is estimated by varying $k$ in a manner based on a
technique of Kitaev \cite{Kitaev95}.
For the present problem, if the number of repetitions of the
$c-\hat{G}$ operator is $k=6$, the measurement result will be
expected as $<\hat{\sigma}_z>_I\,=\,1$.
This implies that before the measurement the control qubit is in
state $|0\rangle$ with a high probability.
However, in practice, operational delays always exist and thus the
wave function of the control qubit acquires a nontrivial dynamical
phase for each delay.
As a consequence, the realistic result of the measurement is
obviously dependent on the {\it total} delay time
$\tau=\sum_{m=1}^k\tau^{(m)}$. We see again that the expected
result is obtained only if the matching condition (14) is
satisfied.

\section{Conclusion and Discussion}

Ideal quantum algorithms usually assume that quantum computing can
be performed by continuously applying a sequence of unitary
transforms. In reality, when performing a practical quantum
computations, there are finite time intervals between the
sequential operations. During these delays, according to the
Schr\"odinger equation, unwanted relative dynamical phases are
acquired by the superposition wave function of the physical qubit
in the quantum register. In general, this phase modifies the
desired quantum interference required for an ideal quantum
computer and thus spoils the correct computational results. Note
that any entanglement between qubits is caused during these
delays, and thus resulting coherent phase errors can be avoided by
simply setting up the total delay times to satisfy certain
matching conditions.
Under these conditions, the relative physical phases in the final
state of the superposition wave function are deleted. Of course,
the dynamical oscillations, due to delays, can also be suppressed
by trivially setting up individual delays
$\tau^{(m)}_j;\,m=1,2...,$ as $\Delta_j\tau^{(m)}_j=2n\pi$. The
key observation here is that \textit{only the total delay time},
instead of the duration for every delay, \textit{needs to be set
up accurately to avoid the coherent dynamical phase errors}.
Therefore, only the proper setting up of the {\it total} delay is
needed for avoiding coherent intrinsic errors. In these
implementations, only the free evolution of the physical qubits in
the index register is considered.

Compared to previous schemes \cite{MSS99, Berman00, Feng01} for
studying similar problems, our scheme presents some advantages.
First, it does not require that the Hamiltonian should be equal to
zero during the quantum register in the idle state (as done in
\cite{MSS99}). Second, operations to force the generation of
additional phases to eliminate these phase errors (as done in
\cite{Berman00}) are not needed. Finally, our approach does not
need to use a pair of degenerate states, formed by using two or
more physical qubits, to encode a logical qubit (as done in
\cite{Feng01}) for transforming the relative phase into a global
phase. Therefore, in principle, our proposal should allow the
implementation of the expected ideal quantum phase estimation
algorithm.

It is worthwhile to emphasize that only the delays between the
sequential functional steps of quantum computing are considered in
the present simplified scheme. The effective dynamical phases,
acquired by superposition wavefunctions of physical qubits during
the effective delays, may be added up, as the key operation
$c-\hat{U}_j$ in the phase estimation algorithm are diagonal in
the logical basis of index register.
The applied non-diagonal Hadamard gate $\hat{H}$ and inverse QFT
operation $\hat{F}^{-1}$ were assumed to be implemented exactly,
and thus the coherent errors relating to the possible operational
delays inside the initialization and measurements had been
neglected. Indeed, the Hadamard gate had been performed exactly by
using one-step operation \cite{Nakamura99}, and the one-step
operational approach had been proposed \cite{Antti03} to exactly
implement the QFT. Furthermore, the present scheme for avoiding
the coherent dynamical phase error is still robust, even if the
operational delays inside the initialization and measurement are
considered. Usually, only a non-diagonal $\sigma_x$-operation is
included in a three-step process for realizing a Hadamard gate,
and in the inverse QFT for measuring a physical qubit.
Fortunately, it is not required to add up the dynamical phase
before and after such a non-diagonal $\sigma_x$-operation in the
quantum phase estimation algorithm. In fact, the qubit is not in
superposition state before (after) the applied
$\sigma_x$-operation in initialization (measurement). Therefore,
in the framework of the quantum phase estimation algorithm, the
present strategy for avoiding the coherent phase error is
sufficiently robust.
This approach can also be used for other quantum algorithms, e.g.,
Deutsch-Jozsa algorithm \cite{Nielsen00}, wherein the key
operation is diagonal and the possible non-diagonal operations are
also only included in the initialization and measurement
operations.

\section*{Acknowledgments}
We acknowledge Drs. Y.X. Liu and X. Hu for discussions. This work
was supported in part by the National Security Agency (NSA) and
Advanced Research and Development Activity (ARDA) under Air Force
Office of Research (AFOSR) contract number F49620-02-1-0334, and
by the National Science Foundation grant No. EIA-0130383.

\end{document}